\documentclass[aps,prl,twocolumn,showpacs,reprint]{revtex4}%
\usepackage{amsfonts}
\usepackage{amsmath}
\usepackage{amssymb}
\usepackage{graphicx}%
\providecommand{\U}[1]{\protect\rule{.1in}{.1in}}

\begin{document}
\title{Tunneling anisotropic magnetoresistance in single-molecule magnet junctions }
\author{Haiqing Xie}
\affiliation{Institute of Theoretical Physics and Department of Physics, Shanxi University,
Taiyuan 030006, China}
\author{Qiang Wang}
\affiliation{Institute of Theoretical Physics and Department of Physics, Shanxi University,
Taiyuan 030006, China}
\author{Hujun Jiao}
\affiliation{Institute of Theoretical Physics and Department of Physics, Shanxi University,
Taiyuan 030006, China}
\author{J.-Q. Liang}

\affiliation{Institute of Theoretical Physics and Department of Physics, Shanxi University,
Taiyuan 030006, China}


\begin{abstract}
We theoretically investigate quantum transport through single-molecule magnet
(SMM) junctions with ferromagnetic and normal-metal leads in the sequential
regime. The current obtained by means of the rate-equation gives rise to the
tunneling anisotropic magnetoresistance (TAMR), which varies with the angle
between the magnetization direction of ferromagnetic lead and the easy axis of
SMM. The angular dependence of TAMR can serve as a probe to determine
experimentally the easy axis of SMM. Moreover, it is demonstrated that both
the magnitude and sign of TAMR are tunable by the bias voltage, suggesting a
promising TAMR based spintronic molecule-device.

\end{abstract}
\date{\today}
\maketitle
\affiliation{Institute of Theoretical Physics and Department of Physics, Shanxi University,
Taiyuan 030006, China}
\affiliation{Institute of Theoretical Physics and Department of Physics, Shanxi University,
Taiyuan 030006, China}
\affiliation{Institute of Theoretical Physics and Department of Physics, Shanxi University,
Taiyuan 030006, China}
\affiliation{Institute of Theoretical Physics and Department of Physics, Shanxi University,
Taiyuan 030006, China}



In the past two decades, magnetoresistance in magnetic tunnel-junctions has
received much attention because of its strong dependence on the relative
magnetization-orientations of two ferromagnetic (FM) electrodes, in terms of
which the tunneling magnetoresistance (TMR) devices are developed for various
applications in magnetic sensors and information storage technology
\cite{Jullier,TMR}. More recently the tunneling anisotropic magnetoresistance
(TAMR) \cite{Gould,Park,Fabian} beyond the conventional TMR has been observed
in the presence of spin-orbit interactions, which depends on the relative
directions between magnetization-orientation of the FM lead and the crystal
axis. Particularly, the TAMR even exists in tunnel junctions with a single
magnetic electrode such as Fe/GaAs/Au \cite{Fabian}, while the conventional
TMR effect does not appear in this configuration. Therefore, the TAMR effect
may lead to new spintronic devices with only one magnetic lead
\cite{Gould,Park,Fabian}.

Single-molecule magnets (SMMs) what we consider possess high spins ($S>1/2$)
and uniaxial magnetic anisotropy with an easy axis, which have potential
applications in molecular spintronics \cite{Wernsdorfer}. In recent
experimental \cite{exp1,exp2} and theoretical
\cite{Kim,Timm2,JBdiode,Kondo,Leuenberger,Martin,JB,JBReview,Rossier,Shen,Xing1,Xing2,Xie,Spinfilter,Renani,JBTMR,JBD,Angle}
studies on electronic transport through magnetic molecules many fascinating
properties have been found, such as complex tunneling spectra \cite{Kim},
negative differential conductance \cite{exp1,Timm2,JBdiode}, Kondo effect
\cite{Kondo}, Berry phase blockade \cite{Leuenberger}, and colossal spin
fluctuations \cite{Martin}. In particular, the spin-polarized transport
through a SMM shows that the magnetization of SMM can be controlled by spin
polarized current \cite{Timm2,JB,JBReview,Rossier}, spin-bias \cite{Shen} and
thermal spin-transfer torque \cite{Xing1}. In addition, a pure spin-current
generated by thermoelectric effects \cite{Xing2}, polarization reversal of
spin-current \cite{Xie}, spin diode behavior \cite{JBdiode} and spin filter
effect \cite{Spinfilter,Renani} are also found. The conventional TMR effect in
spin-dependent transport through a SMM in the sequential, cotunneling and
Kondo regimes is also investigated by Misiorny \textit{et al. }\cite{JBTMR},
where the magnetizations of FM leads are collinear with the magnetic easy axis
of SMM. On the other hand, they also investigate theoretically the magnetic
switching of a SMM coupled to two collinear FM leads, in which the magnetic
easy axis of SMM forms an angle with the FM electrodes \cite{JBD}.

\begin{figure}[ptb]
\centering \vspace{0cm} \hspace{0cm}
\scalebox{0.9}{\includegraphics{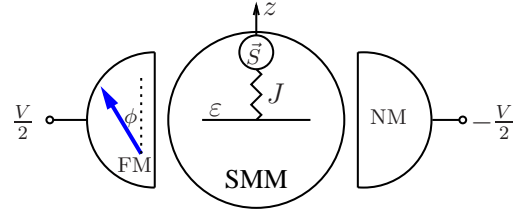}}\caption{(Color online) Schematic of
SMM tunnel junction consisting of FM and normal metallic electrodes. $\phi$
denotes the angle between magnetization of the FM electrode and the easy axis
of the SMM.}%
\label{fig:1}%
\end{figure}

In this paper, we study anisotropic effects of the tunnel junction (see Fig.
\ref{fig:1}), which consists of a SMM sandwiched between one FM electrode and
one normal metallic (NM) electrode, where the FM electrode is not collinear
with the easy axis of SMM different from the TMR case. This setup could be
realized by the scanning tunneling microscope with a spin-polarized tip and a
SMM on a nonmagnetic metallic substrate \cite{JBdiode,SPSTM,Fransson}. We are
going to show that the transport current depends on the magnetization
direction of FM electrode, resulting in the TAMR effect.

The system is described by a Hamiltonian \cite{Timm2,Shen}, $H=H_{leads}%
+H_{SMM}+H_{T}$, in which the first term $H_{leads}=\sum_{\mathbf{\alpha=L,R}%
}\sum_{\mathbf{k,}\tau=\pm}\varepsilon_{\alpha\mathbf{k}}c_{\alpha
\mathbf{k}\tau}^{\dag}c_{\alpha\mathbf{k}\tau}$ is for non-interacting
electrons in the magnetic leads with $c_{\alpha\mathbf{k}\tau}^{\dag}$
($c_{\alpha\mathbf{k}\tau}$) being the creation (annihilation) operator of an
electron in the lead-$\alpha$ (with $\alpha=L,R$ denoting the left and right
leads respectively) of energy $\varepsilon_{\alpha\mathbf{k}}$ and spin index
$\tau=\pm$. In addition, the spin polarization of the FM lead-$\alpha$ is
defined as $P_{\alpha}=(\rho_{\alpha+}-\rho_{\alpha-})/(\rho_{\alpha+}%
+\rho_{\alpha-})$, with $\rho_{\alpha+(-)}$ denoting the state density of the
majority (minority) electrons. The model Hamiltonian of SMM is%
\begin{equation}
H_{SMM}=\sum_{\chi=\pm}\varepsilon d_{\chi}^{\dag}d_{\chi}+Ud_{+}^{\dag}%
d_{+}d_{-}^{\dag}d_{-}-J\mathbf{s}\cdot\mathbf{S}-K(S^{z})^{2}, \label{SMM}%
\end{equation}
where $d_{\chi}^{\dag}$($d_{\chi}$) is the relevant electron creation
(annihilation) operator in the lowest unoccupied molecular orbital (LUMO)
level and $\mathbf{s}\mathbf{\equiv}$ $\sum_{\chi\chi^{\prime}}d_{\chi}^{\dag
}(\mbox{\boldmath$\sigma$}_{\chi\chi^{\prime}}/2)d_{\chi^{\prime}}$ is the
corresponding electron spin operator ($\mbox{\boldmath$\sigma$}$ is the vector
of Pauli matrices, $\chi=\pm$ denotes the spin index). $\varepsilon$ is the
single-electron energy of the LUMO level, which is tunable by the gate
voltage. $U$ represents the Coulomb energy of two electrons with opposite
spins in the LUMO level. $J$ is the exchange coupling parameter between the
spin-$\mathbf{S}$ of SMM and electron spin-$\mathbf{s}$, which can be of
either FM ($J>0$) or antiferromagnetic (AFM) ($J<0$) type, and the parameter
$K>0$ describes the easy-axis anisotropy of SMM. Moreover the total spin is
defined as $\mathbf{S}_{t}\equiv\mathbf{S+s}$\textbf{, }and\textbf{ }the
many-body states of SMM and electron can be written as $\left\vert
n,S_{t};m\right\rangle $, with $n$ denoting the charge state of the molecule,
$S_{t}$ the total spin quantum-number, and $m$ the eigenvalues of
$\mathbf{S}_{t}^{z}$, while the corresponding eigenenegy is $\varepsilon
_{\left\vert n,S_{t};m\right\rangle }$.

The tunneling processes between the molecule and leads can be described by the
Hamiltonian \cite{JBReview,Shen,JBD} $H_{T}=\sum_{\alpha k\chi\tau}t_{\alpha
}c_{\alpha k\tau}^{\dag}(\cos\frac{\phi_{\alpha}}{2}d_{\chi}+\chi\sin
\frac{\phi_{\alpha}}{2}d_{\bar{\chi}})\delta_{\chi\tau}+H.c.$, where
$t_{\alpha}$ denotes the tunnel matrix element of the molecule and the
lead-$\alpha$, and $\phi_{\alpha}$ (with the angle definition $\phi_{L}=\phi$
and $\phi_{R}=0$) is the angle between the magnetization direction of
lead-$\alpha$ and the easy axis of the SMM (as $z$-axis). The spin-dependent
tunnel coupling-strength is described by $\Gamma_{\alpha\pm}=\Gamma_{\alpha
}(1\pm P_{\alpha})/2$ for spin-majority (upper sign) and spin-minority (lower
sign) electrons in the lead-$\alpha$, with $\Gamma_{\alpha}=\Gamma_{\alpha
+}+\Gamma_{\alpha-}$.

For weak coupling between the SMM and leads, i.e. $\Gamma_{\alpha}\ll k_{B}T$
and $\left\vert J\right\vert $, the molecule has enough time to relax to the
eigenstates of $H_{SMM}$ between two consecutive electron tunneling processes
\cite{Shen}. Therefore, we adopt the rate equation approach to study the
stationary transport in the sequential regime. Denoting $P_{i}(t)$ as the
occupation probability of the SMM in the molecular state $\left\vert
i\right\rangle $ at time $t$, we have%
\begin{equation}
\frac{dP_{i}}{dt}=\sum_{\alpha i^{\prime}}W_{\alpha}^{i^{\prime}%
,i}P_{i^{\prime}}-W_{\alpha}^{i,i^{\prime}}P_{i}. \label{Pro}%
\end{equation}
Applying the Fermi's golden rule \cite{Timm2,JBReview}, the sequential
transition rate $W_{\alpha}^{i,i^{\prime}}$ from the state $\left\vert
i\right\rangle $ to $\left\vert i^{\prime}\right\rangle $ with respect to the
lead-$\alpha$ is given by \cite{JBReview,Shen,JBD}%
\begin{align}
W_{\alpha}^{i,i^{\prime}}  &  =\sum_{\chi}\frac{\Gamma_{\alpha}(1+\chi
P_{\alpha})}{2\hbar}[\left\vert \left\langle i\right\vert b_{\chi}\left\vert
i^{\prime}\right\rangle \right\vert ^{2}\nonumber\\
&  f(\varepsilon_{i^{\prime}}-\varepsilon_{i}-\mu_{\alpha})+\left\vert
\left\langle i^{\prime}\right\vert b_{\chi}\left\vert i\right\rangle
\right\vert ^{2}f(\varepsilon_{i^{\prime}}-\varepsilon_{i}+\mu_{\alpha})],
\label{W1}%
\end{align}
where $b_{\chi}=\cos\frac{\phi_{\alpha}}{2}d_{\chi}+\chi\sin\frac{\phi
_{\alpha}}{2}d_{\bar{\chi}}$, $\mu_{\alpha}=(-1)^{\delta_{L\alpha}}eV/2$,
$\varepsilon_{i}$ being the energy of state $\left\vert i\right\rangle $ and
$f(x)$ is the Fermi distribution function. After some algebra, the transition
rate Eq.~(\ref{W1}) can be rearranged as%
\begin{align}
W_{\alpha}^{i,i^{\prime}}  &  =\sum_{\chi}\frac{\Gamma_{\alpha}(1+\chi
P_{\alpha}\cos\phi_{\alpha})}{2\hbar}[\left\vert \left\langle i\right\vert
d_{\chi}\left\vert i^{\prime}\right\rangle \right\vert ^{2}\nonumber\\
&  f(\varepsilon_{i^{\prime}}-\varepsilon_{i}-\mu_{\alpha})+\left\vert
\left\langle i^{\prime}\right\vert d_{\chi}\left\vert i\right\rangle
\right\vert ^{2}f(\varepsilon_{i^{\prime}}-\varepsilon_{i}+\mu_{\alpha})].
\label{W2}%
\end{align}
Furthermore, by introducing the effective tunnel strength $\Gamma_{\alpha,\pm
}=\Gamma_{\alpha}(1\pm P_{\alpha}\cos\phi_{\alpha})/2$ with the effective spin
polarization $P_{\alpha}\cos\phi_{\alpha}$ \cite{Birk}, the magnetization of
FM lead may be considered as collinear with the easy axis of the SMM.

The stationary probabilities obtained from the condition, $\frac{dP_{i}}%
{dt}=0$, with the transition rate together result in the current through
lead-$\alpha$%
\begin{equation}
I_{\alpha}=(-1)^{\delta_{R\alpha}}e\sum_{ii^{\prime}}(n_{i}-n_{i^{\prime}%
})W_{\alpha}^{i,i^{\prime}}P_{i},\nonumber
\end{equation}
and the TAMR ratio in the FM/SMM/NM tunnel junctions is defined as
\cite{Fabian}
\begin{equation}
R_{TAMR}(\phi)=\frac{I(0)-I(\phi)}{I(\phi)}, \label{TAMR}%
\end{equation}
where $I(\phi)$ is the current with the angle $\phi$ between the magnetization
direction of FM lead and the easy axis of the SMM.

\begin{figure}[t]
\includegraphics[width=0.8\columnwidth]{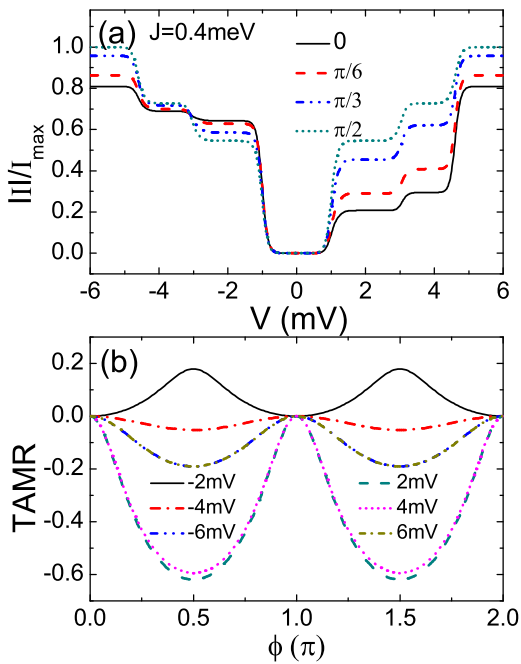}\caption{(Color online) (a)
The absolute value of current as a function of the bias voltage $V$ for
different angles $\phi$, (b) TAMR as a function of the angle $\phi$ for
different bias voltages $V$ with parameters $S=2$, $\varepsilon=0.9$ meV,
$J=0.4$ meV, $U=1$ meV, $K=0.05$ meV, $k_{B}T=0.04$ meV, $P_{L}=0.8$,
$P_{R}=0$, $\Gamma=\Gamma_{L}=\Gamma_{R}=0.001$ meV and $I_{max}=e\Gamma
/\hbar\approx0.25$ nA.}%
\label{fig:2}%
\end{figure}

\begin{figure}[t]
\includegraphics[width=0.8\columnwidth]{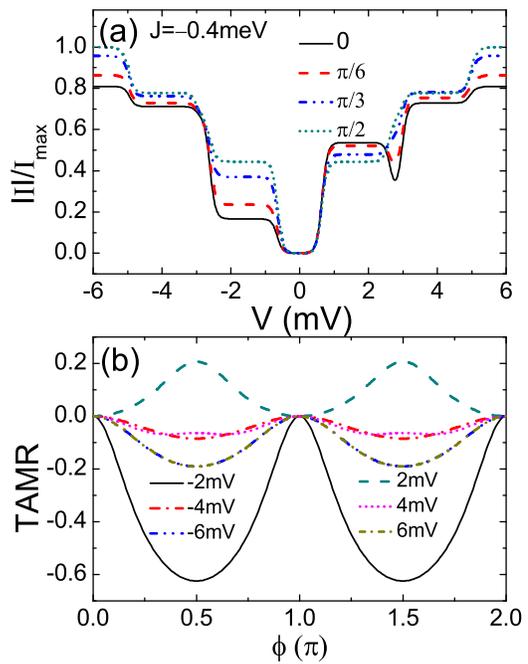} \caption{(Color online) (a)
The absolute value of current as a function of the bias voltage $V$ for
different angles $\phi$, (b) TAMR as a function of the angle $\phi$ for
different bias voltages $V$.}%
\label{fig:3}%
\end{figure}

In Fig. \ref{fig:2}(a), we show the bias-voltage dependence of the current for
different angles $\phi$ with the FM exchange coupling ($J>0$), where only the
angle range $\phi\in\lbrack0,\pi/2]$ is considered, since for $\phi\in
\lbrack\pi/2,\pi]$ one just reverses the polarization direction of FM
electrode. The current spectrum is asymmetric under the bias voltage reversal
due to the asymmetric coupling of the SMM and the electron-spin in the FM
lead\textbf{.} The current magnitude especially in the bias voltage region,
$V=1-5$ mV is more sensitive (with large changes) to the angle variation and
increases with it. The current plateau decreases with the angle increase
around $V=-2$ mV, while the situation is just opposite around the bias voltage
value $V=2$ mV. It is a well known difficulty that the easy axis of SMM
relative to the laboratory frame can not be well determined \cite{Angle} in
SMM transistor experiments \cite{exp1,exp2}, and thus the angular-dependent
current with respect to the FM lead makes it possible to detect experimentally
the magnetic easy axis of SMM \cite{Renani,JBD}.

The TAMR ratio as a function of the angle $\phi$ for different bias voltages
is plotted in Fig. \ref{fig:2}(b), which exhibits a periodic behavior with
increasing the angle $\phi$. The TAMR is positive only for the bias voltage
$V=-2$ mV (solid line), where the current decreases monotonously with the
increase of the angle $\phi$ [see Fig. \ref{fig:2}(a)], and is negative
otherwise. The TAMR amplitudes with $V=2$ mV (dash line) and $V=4$ mV (dotted
line) are larger evidently. Furthermore, when the bias voltage is high enough
all transport channels open up and the TAMR becomes symmetric with respect to
the positive and negative bias voltages $V=\pm6$ mV.

Figure \ref{fig:3} displays transport characteristics for the AFM ($J<0$)
exchange interaction, in which the SMM state $\left\vert 1,3/2;m\right\rangle
$ has a lower energy than the state $\left\vert 1,5/2;m\right\rangle $.
Different from the FM case, the current plateau around $V=\pm2$ mV decreases
(increases) with the increasing angle $\phi$ [Fig. \ref{fig:3}(a)], and the
TAMR for $V=\pm2$ mV is positive (negative) [dash (solid) line in Fig.
\ref{fig:3}(b)]. The TAMR amplitude has the largest value at the bias voltage
$V=-2$ mV and exhibits the same behavior as the FM case for $V=\pm6$ mV.

\begin{figure}[t]
\includegraphics[width=0.8\columnwidth]{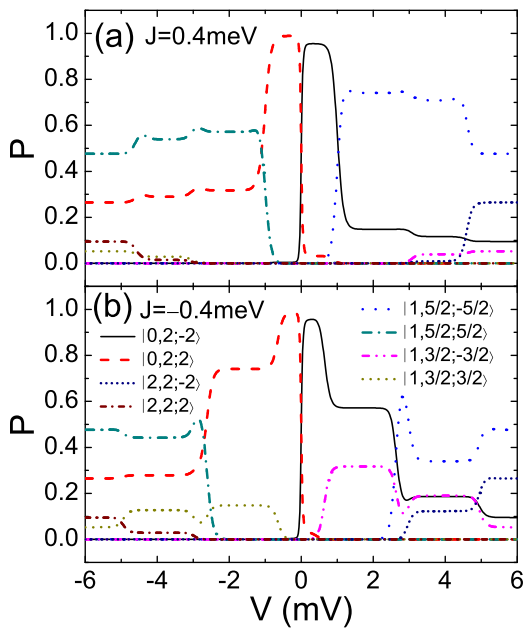}\caption{(Color online) The
main probability distribution of molecular eigenstates as a function of the
bias voltage $V$ for FM (a) and AFM (b) exchange couplings in the collinear
configuration ($\phi=0$). }%
\label{fig:4}%
\end{figure}

In order to understand the characteristic current-spectrum and the TAMR
effect, the main probability distribution of the steady molecular states as a
function of the bias voltage $V$ is plotted in Fig.~\ref{fig:4}, where the FM
electrode is collinear with the easy axis of SMM (i.e. $\phi=0$). We can find
that the steady transport for positive bias voltages is determined by the
states with negative eigenvalues of spin operator $S_{t}^{z}$, while the
situation is just opposite for negative bias voltages. This is because of the
spin-flip process between the transport electron-spin and the SMM
\cite{Timm2,Martin,JBdiode,Rossier}.

In the FM case, the main occupied states of the SMM are $\left\vert
0,2;-2\right\rangle $ and\ $\left\vert 1,5/2;-5/2\right\rangle $ at the bias
voltage $V=2$ mV [see Fig.~\ref{fig:4}(a)], and thus the transport is
dominated by the transition $\left\vert 0,2;-2\right\rangle \Leftrightarrow
\left\vert 1,5/2;-5/2\right\rangle $ of a spin-down electron. The effective
tunnel strength $\Gamma_{L-}=\Gamma_{L}(1-P_{L}\cos\phi)/2$ increases with the
angle $\phi$ variation from $0$ to $\pi/2$, and therefore the current
increases, while the TAMR is negative [dash line in Fig. \ref{fig:2}(b)]. On
the other hand, when $V=-2$ mV, the transport is dominated by the transition
$\left\vert 0,2;2\right\rangle \Leftrightarrow\left\vert
1,5/2;5/2\right\rangle $ of a spin-up electron. Since the effective tunnel
strength $\Gamma_{L+}=\Gamma_{L}(1+P_{L}\cos\phi)/2$ decreases with the
increasing angle $\phi$, the TAMR is positive [solid line in Fig.
\ref{fig:2}(b)]. At the bias voltage $V=4$ mV, the additional main transitions
$\left\vert 0,2;-2\right\rangle \Leftrightarrow\left\vert
1,3/2;-3/2\right\rangle $ (spin-up) and $\left\vert 1,3/2;-3/2\right\rangle
\Leftrightarrow\left\vert 2,2;-2\right\rangle $ (spin-down) take part in the
transport processes, and the TAMR exhibits a similar behavior as the case of
$V=2$ mV. While at $V=-4$ mV the TAMR amplitude is smaller due to the
competition between different transitions of opposite spins.

The state occupations of the AFM exchange interaction are plotted in Fig.
\ref{fig:4}(b), from which we see that the transport is dominated by the
transition $\left\vert 0,2;-2\right\rangle \Leftrightarrow\left\vert
1,3/2;-3/2\right\rangle $ ($\left\vert 0,2;2\right\rangle \Leftrightarrow
\left\vert 1,3/2;3/2\right\rangle $) of a spin-up (spin-down) electron at bias
voltage $V=\pm2$ mV, and thus the TAMR is positive (negative) [dash (solid)
line in Fig. \ref{fig:3}(b)]. The competition between different transitions of
opposite spins suppresses the TAMR amplitudes as shown by dotted and
dash-and-dot lines respectively for $V=\pm4$ mV in Fig. \ref{fig:3}(b).

In summary, the TAMR is obtained explicitly from the angular-dependent
transport in FM/SMM/NM magnetic tunnel junctions by means of the rate-equation
approach. It is demonstrated that the angle between the FM lead and the easy
axis of SMM is crucial to induce the TAMR through the exchange coupling
between the SMM of uniaxial magnetic anisotropy and transport electron-spin.
Both the magnitude and sign of TAMR are efficiently controllable by the bias
voltage, suggesting that this SMM tunnel-junction can be a promising
spintronic device. On the other hand, this TAMR effect may serve as a probe to
detect the uniaxial magnetic anisotropy of the SMM.

This work was supported by National Natural Science Foundation of China (Grant
No. 11075099 and No. 11004124).

\end{document}